\documentclass[twocolumn,nofootinbib,showpacs,prd,superscriptaddress]{revtex4}%
\usepackage{amsmath}
\usepackage{amsfonts}
\usepackage{amssymb}
\usepackage{graphicx,epsf}
\usepackage[titletoc]{appendix}
\usepackage{color}
\usepackage{booktabs}

\newcommand{\tabref}[1]{Tab\,\ref{#1}}

\newcommand{\eqeqref}[1]{Eq.\,\eqref{#1}}
\newcommand{\eqsref}[1]{Eqs.\,\eqref{#1}}
\newcommand{\secref}[1]{Section~\ref{#1}}

\newcommand{\nicenot}[0]{\displaystyle{\not}}
\newcommand{\tr}[0]{\mathrm{tr}}
\newcommand{\Tr}[0]{\mathrm{Tr}}

\newcommand{\mr}[1]{\mathrm{#1}}

\begin{document}

\title{Observed Cosmological Reexpansion in Minimal QFT with Bose and Fermi Fields}


\author{Christine Gruber}
\affiliation{Institut f\"ur Theoretische Physik, Freie Universit\"at Berlin, Arnimallee 14, D-14195 Berlin, Germany;}

\author{Hagen Kleinert}
\affiliation{Institut f\"ur Theoretische Physik, Freie Universit\"at Berlin, Arnimallee 14, D-14195 Berlin, Germany;} 
\affiliation{ICRANet, Piazza della Repubblica 10, 65122 Pescara, Italy.}

\date{\today}

\begin{abstract}
In this work we aim at explaining the re-acceleration of the expansion of the universe, or 
equivalently, the hierarchy problem, with the help of a simple field-theoretical model. In particular, 
we want to account for the notorious discrepancy between the observed value of the cosmological 
expansion term and theoretical values of the vacuum energy of free quantum fields. Rendered infinite 
by short-wavelength fluctuations, evaluation up to a cutoff in momentum space at the Planck 
scale leads to contributions of the order of $10^{76} \,\mr{GeV}^4$. The observed value of the cosmic 
expansion being of the order of $10^{-47} \mr{GeV}^4$, this is a difference of $123$ orders of magnitude. 
We propose a possible resolution of the hierarchy problem by a cancelation of divergences by equivalent 
contributions of bosonic and fermionic fields of the system, albeit after some fine-tuning of the parameters 
of the field theory. We show that in principle nothing beyond conventional ingredients of quantum field theory 
is necessary to provide us with a possible explanation of the observed dark energy, and thus with a solution to 
the hierarchy problem. 

\end{abstract}

\maketitle

\section{Introduction}
Some years ago, astrophysical observations~\cite{SN} 
have suggested that the universe is currently in a transition from a matter-dominated 
state into a phase of accelerating expansion. The cause for this phenomenon has been subject to many speculations, 
reaching from the modification of geometry~\cite{2008Padm,2010Soti,2008Durr}, over the introduction of a cosmological 
constant $\Lambda$ to the postulation of new species of quantum fields and substances like quintessence, k-essence or 
others~\cite{1988Wett,2003Peeb,2006Cope},  
with so far widely unknown properties and origin. The most significant and until now only known fact about such a substance 
is that it has to possess a negative energy density, which would be able to drive the acceleration of expansion. 
Mathematically this corresponds to a constant term in the Friedmann equation, similar to Einstein's cosmological 
constant, or an ideal fluid with negative energy density, i.e., a dependence of pressure on density as 
$p = w \rho$, with $w=-1$. The observed energy density is of the order of 
$\rho_{\Lambda,{\rm{obs}}} = 10^{-47} \mr{GeV}^4$. \\

The standard paradigm of cosmology, known as the $\Lambda$CDM model, features several components to describe the 
observed universe. Besides common baryonic matter, there are two other substances present. One of them is cold 
dark matter (CDM), responsible for the dynamics of compact objects within galaxies and the large-scale structure 
formation of the universe, and the other one is a cosmological constant $\Lambda$, the cause for the expansion of 
the universe. With these ingredients, the equation of state parameter $w$ of the universe can be written as  
\begin{equation}
	w = -\frac{1}{1 + \Omega_M/\Omega_{\Lambda} a^{-3}} \,,
\end{equation} 
where $\Omega_M$ is the normalized density of matter (both baryonic and dark), and $\Omega_{\Lambda}$ 
the normalized density of dark energy in the universe. Observations~\cite{Planck} lead to the estimates 
$\Omega_M \sim 0.3175$ and $\Omega_{\Lambda} \sim 0.6825$, which amounts to a current value of the equation of state 
parameter of the total universe of $w \simeq -0.68$. Alternatively, $w$ can be obtained directly 
from the cosmographic analysis of supernova data~\cite{2012Avil}, which leads to a value of 
$w = -0.7174{_{-0.0964}^{+0.0922}}$, close to the value 
predicted by the $\Lambda$CDM model. This equation of state parameter characterizes the relation between pressure and 
density of the net fluid in the universe and determines the kinematics of its expansion. From the continuity equation, 
and assuming a FLRW-universe parameterized by a scale factor $a(t)$, the density of a fluid evolves as 
\begin{equation}
	\rho \propto a^{-3(1+w)} \,. 
\end{equation}
An equation of state parameter $w = -1$ results in a constant energy density. Solving the first Friedmann 
equation with a constant energy density $\rho$ for the scale factor, this results in 
\begin{equation}
	a(t) \propto e^t \,,
\end{equation}
i.e., an exponential growth of the scale of the universe. \\

Besides the many attempts to model or derive dark energy, an old idea to explain dark energy is to derive it from 
the vacuum energy fluctuations of quantum fields~\cite{2000Sakh}. 
We have to differentiate between bosonic and fermionic contributions to the vacuum energy, which are opposite 
in sign, but else have the same properties. These contributions have the desired behavior in a cosmological context, 
as for example a constant energy density similar to a cosmological constant, and could cause the repulsive effect which 
is necessary to explain the newly accelerated 
expansion of the universe. Unfortunately, however, the predictions for the contributions of quantum 
fields to the vacuum energy are divergent -- or, depending on the method of renormalization applied, 
at least very large. With a momentum cutoff at Planck scale the vacuum energy amounts to a value of 
$\rho_{\Lambda,{\rm{th}}} =10^{76} \,\mr{GeV}^4$. The discrepancy between this theoretical and the observed 
value of the vacuum energy density is thus at least $123$ orders of magnitude, representing the so-called hierarchy problem. \\
The existence of divergences in these expressions is notorious in QFT. Without renormalization schemes, 
the contributions of quantum fields to the vacuum energy are divergent, which would imply that there 
is an infinite energy density filling spacetime. These infinities are generally regarded as an 
unphysical outcome of the methods applied to renormalize the integrals, and usually the paradigm is to 
disregard these divergences, and only consider the energy differences which are detectable in experiments. The 
absolute value should be of no importance and is considered unobservable. This approach is problematic however, 
since there are indeed experiments in which the physical existence of the vacuum energy has been proven. The 
quantization of radiation waves enclosed in a metal box leads to a macroscopic force on metal plates, 
known as the Casimir effect, which is a direct consequence of the existence of the zero-point fluctuations of quantum 
fields. \\ 
On the other hand, astrophysical data tell us that the universe is expanding as if driven by a 
constant, but very tiny, energy density filling the complete space, in behavior much like 
the unphysical divergent vacuum energy contribution from quantum fields. It seems as though 
explaining the universe's dynamics with vacuum fluctuations of fields fails solely due to a mismatch 
of numbers, despite the appealing properties of the vacuum energy. \\
However, there are arguments to validate the identification of those vacuum fluctuations as the energy 
that drives the expansion, like for example by principles of symmetry. Supersymmetry~\cite{SUSY} is a concept 
which predicts a correspondence between the bosonic and fermionic particle content of the universe. Due to 
the fact that the vacuum energies of bosons and fermions have opposite signs, this correspondence can be 
used to eliminate divergences in the vacuum energy by cancelation of opposite-sign terms. In maximally 
supersymmetric models, this balance is exact and leads to a complete elimination of all divergent 
contributions of bosons and fermions to the vacuum energy, whereas in other supersymmetric models divergences 
are canceled only partly. \\
However, recent experiments at LHC have produced results which are inconsistent with 
supersymmetry, or at least with the minimally supersymmetric models amongst them. As an example, the LHCb 
collaboration~\cite{2012Aaij} has reported the decay of $B^0$ into $\mu^+$ and $\mu^-$ at a very low rate, 
which is consistent with the standard model, but expected to be much higher in many of the simpler 
supersymmetric models. In general, these experiments have lead to rather discouraging results for 
supersymmetry~\cite{SUSY-LHC}. Moreover, currently the universe supposedly is in the state of broken 
supersymmetry, which would again lead to an infinite value of the vacuum energy instead of complete 
cancelation of divergences. \\

The approach presented in this work is using a similar argument for the cancelation of divergences, but instead 
of invoking elaborate symmetry principles and complicated mathematical constructions to describe an abundance of 
new particles, it strives for finding the simplest possible balance by asking how many and which kind of particles 
would be needed in addition to the known particles of the standard model in order to yield a vacuum energy that is 
finite and can explain the accelerated expansion of the universe. The goal is to cancel the divergences among 
the quantum fields present in the universe by adding a small amount of new particles by hand, their number and 
nature dictated by 
some simple basic requirements to be fulfilled. By explicitly calculating the vacuum energy for particles 
of different types and masses, it is possible to set up a number of conditions, depending on the masses 
of the particles, which have to be obeyed in order to achieve the correct value of the vacuum energy. 
By appropriately tuning the masses of the fields, it is in principle possible to eliminate the divergences, 
and reduce the vacuum energy to the required tiny amount predicted by astrophysical observations. \\

The paper is organized as follows. In~\secref{sec:basics} we start with some general facts about the 
functional integral formulation of QFT for relativistic bosonic and fermionic fields and the vacuum 
energy of fields. \secref{sec:summingup} presents the exact computations of the vacuum energy and 
sets up conditions for the cancelation of vacuum energy contributions. In~\secref{sec:curving} we 
generalize the calculations to curved spacetimes and obtain the contributions of quantum fields to the vacuum 
energy in terms of an expansion of the effective action with respect to the curvature of spacetime. 
In~\secref{sec:cancelation} we set up the cancelation conditions in curved spacetimes, and evaluate the 
balance for a specific choice of spacetime in~\secref{sec:DE}. It will be shown that 
it is possible, by appropriate choice of masses and fields, to fit the observed dark energy component of the 
universe, and that the occurring contributions of curvature in the cancelation conditions help to simplify 
fulfilling the conditions for the masses of the fields present. \secref{sec:conclusions} contains a conclusion.

\section{Functional integrals of free bosonic and fermionic fields}
\label{sec:basics}
\subsection{Bosonic fields}
In the simplest example of a free complex boson field $\phi$ describing charged bosons with mass $m_{\mr{b}}$ in flat 
spacetime, we can write the relativistic Lagrangian as 
\begin{equation}
 \mathcal{L}_{\rm{b}} = \eta^{\mu\nu} (\partial_\mu \bar{\phi})(\partial_\nu \phi) - m_{\mr{b}}^2 \,\bar{\phi} \phi \,, 
\end{equation}
where $\eta^{\mu\nu}$ is the Minkowski metric. The partition function is given by the functional integral 
\begin{equation}
  Z_{\rm{b}} = \int \mathcal{D}\bar{\phi} \mathcal{D}\phi \, \exp \left[ -i \int\, d^Dx \sqrt{-\eta} \, 
  \,\bar{\phi} \, G_{\rm{b}}^{-1} \, \phi \right] \,,
\end{equation}
where $\eta=-1$ is the determinant of the Minkowski metric. 
With $\partial^2=\eta^{\mu\nu}\partial_{\mu}\partial_{\nu}$, the Greens function or 
propagator in position space is 
\begin{equation}
  G_{\rm{b}}^{-1} = \partial^2 + m_{\mr{b}}^2 \,.
\end{equation}
Carrying out the functional integral with the help of the 
Gaussian integral formula, we obtain 
\begin{equation}
  Z_{\rm{b}} = \det G_{\rm{b}} = \det \left[ \partial^2 + m_{\mr{b}}^2 \right]^{-1} \,.
\end{equation}


\subsection{Fermionic fields}
\label{sec:fermions}
For a free complex massive fermion field $\psi$, the relativistic Lagrangian comes 
from Dirac theory, 
\begin{equation}
  \mathcal{L}_{\rm{f}} = \bar{\psi} (i \nicenot \partial - m_{\mr{f}}\mathbf{1}) \psi \,, 
\end{equation}
where $\nicenot \partial = \gamma_{ij}^{a} \partial_{a}$ is the Feynman-slashed derivative 
operator, and the $\gamma_{ij}^{a}$ are the Dirac $\gamma$--matrices in flat spacetime. 
The functional integral reads 
\begin{equation}
  Z_{\rm{f}} = \int \mathcal{D}\bar{\psi}\, \mathcal{D}\psi \,\exp \left[ i \int\, 
    d^Dx \sqrt{-\eta} \, \, \bar{\psi} \,G_{\rm{f}}^{-1}\, \psi \right] \,. 
\end{equation}
with the inverse propagator 
\begin{equation}
 	G_{\rm{f}}^{-1} = i \nicenot \partial - m_{\mr{f}}\mathbf{1} \,.
\end{equation}
Carrying out the functional integrals over $\psi$ and $\bar{\psi}$, employing a 
slightly different scheme than before due to the Grassmannian nature of fermion fields~\cite{2009Klei}, 
we arrive at 
\begin{equation}
  Z_{\rm{f}} = \det G_{\rm{f}}^{-1} = \det\left[i \nicenot \partial - m_{\mr{f}}\mathbf{1}\right] \,,
\end{equation}
Arguing that, due to the existence of the Dirac sea, the energies of Dirac fermions are 
symmetric around $p=0$, and considering that for an $n \times n$--matrix with even $n$, an overall sign 
in the determinant does not matter, we can rewrite the determinant in the partition function as 
\begin{equation}
 Z_{\rm{f}} = \det[\nicenot \partial^2 + m_{\mr{f}}^2\mathbf{1}]^{1/2} \,.  
\end{equation}

\subsection{Effective action}
\label{sec:tracelogs}
Combining the results for the partition function for free bosons and fermions, and 
introducing the notion of effective action $S_{\mr{eff}}$ as 
\begin{equation}
  Z = e^{i S_{\mr{eff}}} \,,
\end{equation}
we can express the first loop contribution to the effective action for bosons as 
\begin{equation}
  i S_{\mr{eff}}^{(\rm{b})} = \ln \det G_{\rm{b}}  
  		=  - \Tr_{\rm{F}} \ln\left(\partial^2 + m_{\mr{b}}^2\right) \,,
\end{equation}
where $\Tr_{\rm{F}}$ denotes the trace over the functional space. \\
In the case of fermions, we have 
\begin{equation}
  i S_{\mr{eff}}^{(\rm{f})} = - \ln \det G_{\rm{f}} = 
    \frac{1}{2} \, \Tr_{\rm{F,D}} \ln \left(\nicenot \partial^2 + m_{\mr{f}}^2\mathbf{1}\right) \,,
\end{equation}
where the subscript $D$ indicates the trace over the Dirac indices, which leads to an additional factor 
of four in the expression. Carrying out the Dirac trace and summarizing the results, we can write 
the partition function of a system with charged bosons and fermions as 
\begin{equation}
  Z = \exp\left[ - \Tr_{\rm{F}} \ln\left(\partial^2 + m_{\mr{b}}^2\right) + 
    2 \,\Tr_{\rm{F}} \ln \left(\partial^2 + m_{\mr{f}}^2\right) \right] \,.
\end{equation}
We see that the contributions of bosons and fermions enter with opposite sign and 
differ from each other by a factor of two. This mirrors the fact that complex scalar fields 
possess only two degrees of freedom (for charge conjugation), whereas massive Dirac fermions have 
four degrees of freedom (two for the charge conjugation and two for the spin orientations). 
Considering variations as a real scalar field, i.e., uncharged bosons, or a Majorana spinor, 
i.e., uncharged fermions, the partition functions would differ in the 
prefactor of the exponent's argument, according to the respective number of degrees of 
freedom. This holds for other species like vector particles as well, which have 
additional numerical factors due to the possible spin orientations and charges. In the example 
of a vector boson, its degrees of freedom can be inferred from the formulation as a 
superposition of plane waves with different polarization states $\epsilon^{\mu}(\nu)$. For a 
massive boson, three polarization modes are possible, i.e., two transverse and one longitudinal 
mode, whereas for a massless boson, the longitudinal mode vanishes, and thus two degrees of 
freedom are left. A charge signifies further multiplications of the degrees of freedom. \\
The general rule is that bosons and fermions yield contributions to the vacuum energy which are 
opposite in sign, and this fact will be used to set up a cancelation scheme. Bosons occur in the 
effective action with a negative sign, corresponding to a set of harmonic oscillators with positive 
energies, whereas fermions have negative energies and contribute positive terms to the effective 
action. \\
The one-loop contribution to the effective action, as thus formulated for bosons and fermions, 
is the quantity of interest to use for the balance of bosonic and fermionic contributions. 
Hereby we neglect any form of interaction between the particle species and consider purely 
free fields. To be more realistic, one would have to consider all possible further diagrams, including 
higher order self-interactions, but for the purpose of demonstrating the principle of our model, we 
restrict ourselves to the first loop diagrams.

\section{Calculating the total vacuum energy}
\label{sec:summingup}
In order to calculate the vacuum energy of free particles described by the Lagrangians 
introduced before, we transfer the problem to Fourier space, where we are left to carry out 
integrals of the form 
\begin{equation} \label{eq:integral}
  i S_{\rm{eff}} \propto \int d^Dx \sqrt{-\eta} \,\int \frac{d^{D}p}{(2\pi)^D}\, 
    \ln\left(-p^2+m^2\right) \,. 
\end{equation}
Integrals of that kind for the case of $D=4$ are divergent, but can be calculated using regularization techniques. 
A commonly used method is cutoff regularization, where the integral in phase space is carried out 
only up to a maximum frequency, which represents the divergence. Another option is to use 
dimensional regularization, which will be employed here, and which substitutes the number of spacetime 
dimensions $D$ by $D=d-\epsilon$, $d$ being an integer. This will transform the divergences into 
the form of poles in $\epsilon$ in the limit of physical dimensions, i.e., $\epsilon \rightarrow 0$. 
Whichever regularization scheme is used, there will be divergences in the resulting expression, and depending on 
the regularization they will take different shapes. Using dimensional regularization, 
for a general number of spacetime dimensions $D$, we can calculate the integral in~\eqeqref{eq:integral} 
as (see e.g.~\cite{2001HKVSFp107})
\begin{equation}
  \int{\frac{d^Dp}{(2\pi)^D}\, \ln\left(-p^2+m^2\right)} = 
    -i \,\frac{\Gamma(-D/2)}{(4\pi)^{D/2}} \, \frac{1}{(m^2)^{-D/2}}. 
\label{eq:HKVSF}
\end{equation}
This integral is divergent because the $\Gamma$--function diverges for negative even 
integer values without any possibility of analytic continuation or similar methods. 
The expression can be processed further using aforementioned substitution 
$D=4-\epsilon$, which will make it possible to explicitly isolate 
the infinities in terms which diverge in the limit $\epsilon \rightarrow 0$.
We use the expansion of the $\Gamma$--function for small $\epsilon$~\cite{2001HKVSFp128}, 
\begin{eqnarray} \label{eq:Gammaexp}
    \Gamma(-n+\epsilon) &\simeq& \frac{(-1)^n}{n!} \bigg\{ \frac{1}{\epsilon} + 
	\psi (n+1) \\
    &~& + \frac{\epsilon}{2} \left[ \frac{\pi^2}{3} + \psi^2 (n+1)
	- \psi' (n+1) \right] \bigg\} \,, \nonumber
\end{eqnarray}
where $\psi(n)$ is the Digamma function, and $\psi'(n)$ the Trigamma function, 
and expand the mass term for small $\epsilon$ as well, 
\begin{equation} \label{eq:massexp}
  m^{-\epsilon} = \mu^{-\epsilon}\, \left( \frac{\mu}{m} \right)^{\epsilon} \simeq 
    \mu^{-\epsilon}\, \left[ 1+\epsilon \,\ln \frac{\mu}{m} + \mathcal{O}(\epsilon^2)\right] \,.
\end{equation}
Here $\mu$ is an auxiliary parameter with the dimension of a mass and of arbitrary size, introduced to 
make the argument of the logarithm dimensionless. \\
Truncating at linear order in $\epsilon$, we end up with an expression for the 
effective action as 
\begin{equation}
  S_{\rm{eff}} \simeq -\frac{i\mu^{-\epsilon}}{(4\pi)^{2}} \, m^4 \,
    \left[ \frac{1}{\epsilon} + \ln \frac{\mu}{m} + 1 -\frac{\gamma}{2} + 
    \mathcal{O}(\epsilon)\right] \,,
\end{equation}
where $\gamma$ is the Euler--Mascheroni constant. 
From this result, we see that in order to cancel the divergent terms proportional to 
$1/\epsilon$, we need to require that the sum over all quartic powers of the 
masses in the system vanish, 
\begin{equation}
  \sum_{\rm{b}} m_{\rm{b}}^4 = \sum_{\rm{f}} m_{\rm{f}}^4 \,, 
  \label{eq:firstcond}
\end{equation}
which will thus ensure that the effective energy is finite. This will also cancel 
the constant terms, $1 -\gamma/2$, in the effective action. The remaining convergent 
terms in the effective action should subsequently be tuned to obtain the observed cosmological 
constant, which can be achieved by fulfilling the condition 
\begin{equation} \label{eq:secondcond}
  \sum_{\rm{b}} m_{\rm{b}}^4 \,\ln \frac{\mu}{m_{\rm{b}}} - 
    \sum_{\rm{f}} m_{\rm{f}}^4 \,\ln \frac{\mu}{m_{\rm{f}}} = \rho_{\Lambda}\,.
\end{equation}
In this derivation, we have failed to obtain an additional divergence, since 
within the formalism of dimensional regularization quadratically divergent 
contributions to the integral are lost due to the application of Veltman's 
rule~\cite{1975Leib,2001HKVSFp107}. We would have obtained those divergences 
correctly if we had used a cutoff regularization, which would have lead to an additional 
term in the effective action proportional to the cutoff scale, but without any mass 
dependence. In order to reconsider those divergences in the balance, we additionally have to 
require the fulfillment of another condition, i.e., the balance of the degrees of 
freedom $\nu_i$ of the particles present in the system, such that the contribution 
proportional to the cutoff scale will be canceled. This is achieved by the 
condition 
\begin{equation} \label{eq:dof}
  \sum_{\rm{b}} \nu_{\rm{b}} = \sum_{\rm{f}} \nu_{\rm{f}} \,. 
\end{equation}
With these three conditions,~\eqeqref{eq:firstcond},~\eqref{eq:secondcond} and~\eqref{eq:dof}, 
we can make the divergent contributions to the vacuum energy vanish and obtain the correct 
remainder to explain the accelerated expansion of the universe.

\section{Results in curved spacetime}
\label{sec:curving}
Before we continue to investigate how these conditions can be fulfilled, we move on to reconsider 
the situation in curved spacetime described by a general metric $g^{\mu\nu}$, where the 
occurring operators are replaced by generalized expressions in curved spacetime. \\
The Laplacian can be written in a conformally invariant way as the so-called generalized 
Laplace--Beltrami operator, 
\begin{equation}
 \Delta^{\mr{gen}}_{LB} = \Delta_{LB} - \xi R = \frac{1}{\sqrt{-g}}\partial_{\mu} \sqrt{-g} 
   g^{\mu\nu} \partial_{\nu} - \xi R. 
\end{equation}
where $R$ is the curvature scalar of spacetime and $\Delta_{LB}$ is the usual Laplace operator 
in curved spaces. It can be shown~\cite{2009Kleip802} that the factor $\xi$ must be zero, on 
grounds of demanding a unique mapping between the coordinates of flat and curved spacetimes, not only 
for point particles, but also for extended objects of different shape like wave packets. \\
For fermionic fields, the Dirac $\gamma$--matrices in flat spacetime $\gamma_{ij}^{a}$ have to be replaced by 
\begin{equation}
  \gamma_{ij}^{\mu} = e_{a}^{\mu} \gamma_{ij}^{a} \,,
\end{equation}
where $e_{a}^{\mu}$ are the vierbein fields corresponding to the geometric description of the spacetime
$g^{\mu\nu}$. Furthermore, the factor $\sqrt{-\eta}$ in the action is not equal to unity anymore, 
but has to be replaced by the determinant of the metric of curved spacetime, $\sqrt{-g}$. \\
Quantitatively the effects of curvature can be described in terms of a power series expansion, 
which will be introduced and discussed in detail in the next subsections.

\subsection{Curvature expansion for scalars}
The effects of curvature on the vacuum energy of fields can be calculated within the framework 
of a so-called Schwinger--de Witt expansion, in which the Greens function of the theory is 
expanded into a power series with coefficients depending on the curvature of spacetime. The theory 
of quantum fields in curved spacetimes in general and the calculation of Greens functions in 
particular have been first introduced by de Witt~\cite{1965deWi}, and extensively treated in several 
works by Christensen, Bunch, Parker, Toms and 
Vassilevich~\cite{1976Chri,1978Chri,1979Bunc,1985Park,1985Par1,2003Vass}. To apply the formalism 
to our present purpose, we have to express the effective action, i.e., the quantitative measure for 
the vacuum energy of the universe, in terms of the respective Greens functions of the system. \\
To this end, by introducing an integral over $m^2$ we can rewrite the effective action for a scalar 
field, denoted by the superscript $(0)$, as
\begin{equation}
  iS^{(0)}_{\mr{eff}} = -\int d^Dx \sqrt{-g} \, \int dm^2 \frac{1}{\partial^2 + m^2} \,, \nonumber
\end{equation}
where we can then identify the propagator in position space as 
\begin{equation}
  G(x,x') = \frac{1}{\partial^2 + m^2} \,,
\end{equation}
fulfilling the equation 
\begin{equation}
  \left( \partial^2 + m^2 \right) G(x,x') = -\delta(x,x')\,.
  \label{eq:PropEoM}
\end{equation}
Following~\cite{1965deWi,1978Chri}, we define the operator acting on $G(x,x')$ on the left hand 
side as $\hat{F}$. The propagator is now written as an integral over the heat kernel 
$\langle x,s|x',0 \rangle$ as 
\begin{equation}
  G(x,x') = -i \int_0^{\infty} \langle x,s|x',0 \rangle e^{-im^2 s} ds \,,
 \label{eq:propagator}
\end{equation}
where $s$ is the pseudotime. According to the formalism of the Schwinger--de Witt expansion, the 
heat kernel can be further decomposed to 
\begin{equation}
 \langle x,s|x',0 \rangle = \frac{i \Delta_{VM}(x,x')}{(4\pi i s)^{D/2}} \,
    e^{i \sigma/2s}\, \Omega(x,x',s) \,,
 \label{eq:kernel}
\end{equation}
where $\Omega(x,x',s)$ is a newly introduced function to be determined, 
$\Delta_{VM}(x,x')$ is the van Vleck--Morette determinant, and 
$\sigma = g_{\mu\nu}\sigma^{\mu}\sigma^{\nu}$ is the geodesic difference 
between the points $x$ and $x'$. \\
The determining equation for $\Omega(x,x',s)$ can be obtained by plugging 
the expansion for the kernel,~\eqeqref{eq:kernel}, into the 
equation for the propagator,~\eqeqref{eq:PropEoM}, leading to 
\begin{equation}
 i \frac{\partial}{\partial s} \langle x,s|x',0 \rangle = \hat{F} \big|_{m=0} \langle x,s|x',0 \rangle\,,
\end{equation}
with the operator $\hat{F}$ evaluated for $m=0$. Subsequently, the 
equation for the function $\Omega(x,x',s)$ can be found as 
\begin{equation}
 i \frac{\partial}{\partial s} \Omega + \frac{i}{s} \Omega^{;\mu} \sigma_{;\mu} = 
  -\Delta^{-1/2} (\Delta^{1/2} \Omega)_{;\mu}^{~~\,\mu} \,.
\label{eq:Omega}
\end{equation}
To solve this equation we write $\Omega$ as a power series in the pseudotime $s$, 
\begin{equation}
 \Omega(x,x',s) = \sum_{j=0}^{\infty}(is)^j a_j(x,x')\,,
\end{equation}
where the $a_j$ are determined by the recursion relations which 
are obtained by plugging this ansatz back into~\eqeqref{eq:Omega}: 
\begin{eqnarray}
 \sigma^{;\mu} a_{0;\mu} &=&0\,,\\
 \sigma^{;\mu} a_{j+1;\mu} + (j+1) a_{j+1} &=& \Delta^{-1/2}\, 
      \hat{F}\big|_{m=0} \left[ \Delta^{1/2}\, a_j \right] \,,\nonumber
\end{eqnarray}
with the boundary condition $a_0(x,x') = 1$. This boundary condition ensures the consistency 
of the curvature expansion with the limit of flat spacetimes, in which all higher order 
coefficients with $j\geq1$ will vanish, and only the $0$th order coefficient remains. 
In the coincidence limit $x\rightarrow x'$, the van Vleck--Morette determinant becomes the unit 
matrix, and the geodesic distance $\sigma=g_{\mu\nu}\sigma^{\mu}\sigma^{\nu}$ between $x$ and $x'$ 
becomes zero. The heat kernel then reads  
\begin{equation}
 \langle x,s|x,0 \rangle = \frac{i}{(4\pi i s)^{D/2}} \sum_{j=0}^{\infty}(is)^j a_j(x,x) \,,
 \label{eq:kernel2}
\end{equation}
with the first three coefficients obtained from the above recursion relations as 
\begin{subequations} \label{eq:expParker}
\begin{align}
 a_0 &= 1 \,, \label{eq:expParker0} \\
 a_1 &= \frac{1}{6} R\,, \label{eq:expParker1} \\
 a_2 &= \frac{1}{30} \square R - \frac{1}{72} R^2 + \frac{1}{180} \left( R_{\alpha \beta \gamma \delta} 
      R^{\alpha \beta \gamma \delta} - R_{\alpha \beta} R^{\alpha \beta} \right) \,. \label{eq:expParker2} 
\end{align}
\end{subequations}
After introducing the identity 
\begin{equation}
 \frac{i}{(4\pi is)^{D/2}} \,e^{-im^2s} = \int \frac{d^Dk}{(2\pi)^D} \,e^{-is(-k^2+m^2)} \,,
\end{equation}
we replace the $j$th power of pseudotime $(is)^j$ with the $j$th derivative 
with respect to $m^2$.  Carrying out the integral over the pseudotime $s$ and plugging 
everything back into the propagator, we obtain 
\begin{equation}
 G(x,x) =\int\frac{d^Dk}{(2\pi)^D}\,{\sum_{j=0}^{\infty}a_j(x,x) 
      \left( -\frac{\partial}{\partial m^2} \right)^j \, \left[\frac{1}{-k^2+m^2}\right]} \,,
\end{equation}
and thus for the effective action 
\begin{widetext}
\begin{equation}
 iS^{(0)}_{\mr{eff}} = -\int{d^Dx \sqrt{-g} \,\,\int{dm^2} \int\frac{d^Dk}{(2\pi)^D}\, \left( \frac{a_0}{-k^2+m^2} + 
    \frac{a_1}{(-k^2+m^2)^2} + \frac{2 a_2}{(-k^2+m^2)^3} + ... \right) } \,.
  \label{eq:effaction}
\end{equation}
\end{widetext}
We have thus obtained the effective action in terms of a series expansion in $(-k^2+m^2)^{-1}$, with 
coefficients depending on the curvature of spacetime. What remains to be done is carrying out the 
integrals over $k$ and $m^2$. The divergent $k$-integrals have to be calculated by a renormalization 
technique, similar to the divergent integrals in the case of flat spacetime treated earlier, whereas 
the mass integrals are straightforward. First however we will derive the curvature expansion for other 
species of particles, in particular spinor and vector fields.

\subsection{Curvature expansion for spinors}
The effective action for a spinor field, denoted by the superscript $(1/2)$, is 
\begin{equation}
 	iS^{(1/2)}_{\mr{eff}} = - \int d^Dx \sqrt{-g} \, \tr_{\rm{D}} \, \ln
 		\left( i\nicenot\partial + m\mathbf{1} \right) \,, 
\end{equation}
where $\tr_{\rm{D}}$ represents the trace over the Dirac indices of the spinor field. 
Introducing the integral over the mass, we can rewrite the expression as 
\begin{equation}
 	iS^{(1/2)}_{\mr{eff}} = - \int d^Dx \sqrt{-g} \, \int dm \,\tr_{\rm{D}} 
 		\left( \frac{1}{i\nicenot\partial + m\mathbf{1}} \right) \,,
\end{equation}
and identify the propagator in position space as 
\begin{equation}
 G(x,x') = \frac{1}{i\nicenot\partial + m\mathbf{1}} \,,
\end{equation}
fulfilling the equation 
\begin{equation}
 \left( i\nicenot\partial + m\mathbf{1}\right) G(x,x') = -\delta(x,x')\,.
\end{equation}
However, to be able to apply the Schwinger--de Witt expansion, we need a 
quadratic operator, not one proportional to $\partial$ as in the case of Dirac spinors. 
Thus, we rewrite the effective action as outlined in~\secref{sec:fermions} to 
\begin{equation}
	iS^{(1/2)}_{\rm{eff}} = \frac{1}{2} \int d^Dx \sqrt{-g} \, \,\tr_{\rm{D}} \ln \left[ \nicenot \partial^2 + m^2\mathbf{1} \right] \,,
\end{equation}
now containing a quadratic operator, which is taken into account by the 
factor of $1/2$ in front of the whole expression. Introducing now the mass integral 
as before in the case of the scalar field, the effective action becomes 
\begin{equation}
	iS^{(1/2)}_{\rm{eff}} = \frac{1}{2} \int d^Dx \sqrt{-g} \, \int dm^2 \,\tr_{\rm{D}} \,\frac{1}{\nicenot \partial^2 + m^2\mathbf{1}} \,,
\end{equation}
where now we can carry out the curvature expansion as before for the Greens 
function $\mathbf{G}(x,x')$ obeying the equation 
\begin{equation}
	\tr_{\rm{D}} \left(\nicenot\partial^2 + m^2\mathbf{1} \right) \mathbf{G}(x,x') = 
	  -\delta(x,x') \,. 
\end{equation}
Taking the trace, this equation can be brought into the form 
\begin{equation}
	  \left( \partial^2 + \frac{1}{4} R - m^2 \right) G(x,x') = 
	  -\delta(x,x') \,. 
\end{equation}
The Greens function can then be written in terms of the heat kernel expansion 
as 
\begin{equation} \label{eq:SpinorGreensfunction}
 	\mathbf{G}(x,x) = \int\frac{d^Dk}{(2\pi)^D}\,{\sum_{j=0}^{\infty} \mathbf{a}_j(x,x) 
      \left( -\frac{\partial}{\partial m^2} \right)^j \, \frac{1}{-k^2+m^2}} \,,
\end{equation}
with the first coefficients for spinor fields as~\cite{1978Chri}
\begin{subequations} \label{eq:coeffSpinors}
\begin{align}
  \mathbf{a}_0 &= \mathbf{1} \,, \label{eq:coeffSpinors0} \\
  \mathbf{a}_1 &= \frac{1}{12} R \mathbf{1} \,, \label{eq:coeffSpinors1} \\
  \mathbf{a}_2 &=  \big( \frac{1}{288} R^2 -\frac{1}{120} \square R - 
  		\frac{1}{180} R_{\alpha \beta} R^{\alpha \beta} \nonumber \\
     &~+ \frac{1}{180} R_{\alpha \beta \gamma \delta} R^{\alpha \beta \gamma \delta} \big) \mathbf{1}
     - \frac{1}{192} \sigma_{\alpha\beta} \sigma_{\gamma\delta} R^{\alpha \beta \lambda \xi}_{~~} 
	R^{\gamma \delta}_{~\,~\lambda \xi} \,, \label{eq:coeffSpinors2} 
\end{align}
\end{subequations}
and $\sigma_{\alpha\beta} = \frac{i}{2} \bigl[ \gamma_{\alpha},\gamma_{\beta} \bigr]_-$ 
being the commutator of the $\gamma$--matrices. Considering that in the spinor case we 
still have to evaluate the Dirac trace over the coefficients, the effective action is 
then obtained as 
\begin{equation}
 iS^{(1/2)}_{\rm{eff}} = \frac{1}{2} \int d^Dx \sqrt{-g} \, \int dm^2 \,\tr_{\rm{D}} \,\mathbf{G}(x,x) \,,
\end{equation}
with $\mathbf{G}(x,x)$ determined by~\eqsref{eq:SpinorGreensfunction} and~\eqref{eq:coeffSpinors}.

\subsection{Curvature expansion for vector fields}
Next we investigate the expansion for a vector field included in 
the balance. We start with the action for a massive vector field, 
denoted by the superscript $(1)$, 
\begin{eqnarray}
	S^{(1)} &=& \frac{1}{2} \int d^Dx \sqrt{-g} \, A_{\mu} \bigg[ g^{\mu \nu} (\partial^2 
		+m^2) \\
	&& ~~~~~~~~~~~~~~~~~~~~~~~~~~- \left( 1-\frac{1}{\alpha} \right) \partial^{\mu} 
		\partial^{\nu} \bigg] A_{\nu} \,. \nonumber
\end{eqnarray}
We include a gauge-fixing term to account also for the possibility of taking 
the limit $m\rightarrow 0$ later, which considers the case of photons or gluons. 
The operator in the action is identified as the inverse of the 
Greens function, 
\begin{equation}
 \left(G^{\mu\nu}\right)^{-1} (x,x') = g^{\mu\nu} (\partial^2 + m^2) - 
      \left( 1-\frac{1}{\alpha} \right) \partial^{\mu} \partial^{\nu} \,,
\end{equation}
and obeys the equation 
\begin{equation}
 \left[ g^{\mu\nu}(\partial^2+m^2) - \left( 1-\frac{1}{\alpha} \right)\partial^{\mu} 
      \partial^{\nu} \right] G_{\mu\nu}(x,x') = -\delta(x,x') \,.
  \label{eq:vectorEqu}
\end{equation}
The effective action then reads 
\begin{equation}
 iS^{(1)}_{\rm{eff}} = - \int d^Dx \sqrt{-g} \, \tr_{\rm{L}} \, \ln \, \left(G^{\mu\nu}\right)^{-1} (x,x')  \,,
\end{equation}
where $\tr_{\rm{L}}$ denotes the trace over the Lorentz indices. 
The introduction of a mass integral can be done considering the rules of 
matrix computation and logarithms, and leads to 
\begin{equation}
 iS^{(1)}_{\rm{eff}} = - \int d^Dx \sqrt{-g} \int dm^2 \, \tr_{\rm{L}} \left[ \, g^{\mu \lambda} G_{\lambda \nu}(x,x') \right] \,.
\end{equation}
The matrix $G_{\lambda \nu} (x,x')$ can be expanded and calculated in the same way as 
the propagators before in the framework of a Schwinger--de Witt expansion. The 
case of vector fields has been extensively treated in~\cite{1978Chri}, where 
the first three coefficients in the coincidence limit were found to be 
\begin{widetext}
\begin{subequations} \label{eq:coeffVectors}
\begin{align}
   a_{0\lambda \nu}^{~} &= \delta_{\lambda \nu} \,, \label{eq:coeffVectors0} \\
   a_{1\lambda \nu}^{~} &= \frac{1}{6}\left( R g_{\lambda \nu} - R_{\lambda \nu} \right) 
  		\,, \label{eq:coeffVectors1} \\
   a_{2\lambda \nu}^{~} &=  \Biggl[ -\frac{1}{6} R \, R_{\lambda \nu} -\frac{1}{6} \square R_{\lambda \nu} + 
    \frac{1}{2} R^{}_{\lambda \alpha} R^{\alpha}_{~\nu} - 
    \frac{1}{12} R^{\alpha\beta\gamma}_{~~~~\nu} R^{}_{\alpha\beta\gamma\lambda} \nonumber\\
   &~~~~~~~~+ \left( \frac{1}{72} R^2 +\frac{1}{30} \square R - 
    \frac{1}{180} R_{\alpha \beta} R^{\alpha \beta}
    + \frac{1}{180} R_{\alpha \beta \gamma \delta} R^{\alpha \beta \gamma \delta}
    \right) \, g_{\lambda \nu} \Biggr] \,.
    \label{eq:coeffVectors2} 
\end{align}
\end{subequations}
\end{widetext}
Note that these results have been obtained using the Feynman gauge, i.e., $\alpha=1$. 
Within the effective action, we then have to consider an additional contraction with 
the metric, i.e., calculate the expressions $g^{\mu\lambda} a_{j\lambda \nu}$, and then take 
the trace over the Lorentz indices $\mu,\nu$.

\section{Balancing the effective action}
\label{sec:cancelation}
To obtain the final expressions for the effective actions, we investigate the 
obtained expressions limiting the calculations to the first three terms in the expansion 
of the Greens function, i.e., $j\leq 2$. We have to solve integrals of the form 
\begin{equation}
 I_{\alpha}(D) = \int \frac{d^Dk}{(2\pi)^D} \, \frac{1}{(-k^2+m^2)^{\alpha}} \,,
\end{equation}
for $\alpha=1,2,3$. Similarly to before, the results can be expressed in terms of 
the Gamma function, 
\begin{equation}
 I_{\alpha}(D) = \frac{i}{(4\pi)^{D/2}} \, \frac{\Gamma\left(\alpha-\frac{D}{2}\right)}{\Gamma(\alpha)} 
    \, \frac{1}{(m^2)^{\alpha-D/2}} \,.
\end{equation}
We use dimensional regularization to further process those integrals, setting $D=4-\epsilon$, 
with $\epsilon$ taken to zero in the end. In further calculations, 'safe' limits $\epsilon 
\rightarrow 0$ will be taken immediately. The integral with $\alpha=3$ is convergent and can be 
expressed directly, whereas the cases $\alpha=1,2$ contain divergences which have to be investigated: 
\begin{eqnarray}
 I_{1}(4-\epsilon) &=& \frac{i}{(4\pi)^{2}} \, \Gamma\left(\frac{\epsilon}{2}-1\right) \, \left(m^2\right)^{1-\epsilon/2} \,,\\
 I_{2}(4-\epsilon) &=& \frac{i}{(4\pi)^{2}} \, \Gamma\left(\frac{\epsilon}{2}\right) \, \left(m^2\right)^{-\epsilon/2} \,,\\
 I_{3}(4-\epsilon) &=& \frac{i}{2\,(4\pi)^{2}} \, \frac{1}{m^2} \,.
\end{eqnarray}
Before considering the divergences however, we carry out the mass integrations, 
which in the case of the scalar field leads to an effective action of 
\begin{widetext}
\begin{equation} \label{eq:scalars}
  S_{\rm{eff}}^{(0)} = -\frac{1}{8\pi^2} \int d^{4-\epsilon}x \sqrt{-g} \,
      \Big[ a_0\, \Gamma\left(\frac{\epsilon}{2}-1\right)
    	\, \frac{m^{4-\epsilon}}{4-\epsilon}  + \,a_1 \,\Gamma\left(\frac{\epsilon}{2}\right) \, 
      \frac{m^{2-\epsilon}}{2-\epsilon} + a_2 \,\ln m - 
      ...\Big]\,, 
\end{equation}
\end{widetext}
with the coefficients $a_j$ given by~\eqsref{eq:expParker}. 
For the spinor field, denoted by the superscript ${(1/2)}$, integrating out the 
mass is again straightforward. After taking the Dirac trace, we can write 
the effective action for spinor fields in the same form as for scalars, but with 
the opposite sign, and 
with the new coefficients $\tilde{a_i}$, 
\begin{subequations} \label{eq:spinorcoeff}
\begin{align}
 	& \tilde{a}_0 = 2 \,, \\
 	& \tilde{a}_1 = \frac{1}{6} R \,, \\ 
	& \tilde{a}_2 = 
		\frac{1}{144} R^2 -\frac{1}{60} \square R - \frac{1}{90} R_{\alpha \beta} R^{\alpha \beta}
      + \frac{1}{90} R_{\alpha \beta \gamma \delta} R^{\alpha \beta \gamma \delta} \nonumber\\ 
      &~~~~ +\frac{1}{96} \tr_{\rm{D}} \, \left[ \sigma_{\alpha\beta} \sigma_{\gamma\delta} R^{\alpha \beta \lambda \xi}_{}
		R^{\gamma \delta}_{~\,~\lambda \xi} \right]\,.
\end{align}
\end{subequations}
In the third case of the vector bosons, the effective action is basically the same as 
in the scalar boson case, including the sign, but with coefficients where the trace over 
the Lorentz indices still has to be taken. \\ 
As noted earlier, the effective action for bosons is negative, corresponding to positive physical 
energies, whereas for fermions the effective action is positive, implying negative energies.
In order to obtain a reasonable result in the limit $\epsilon \rightarrow 0$, we can now expand 
the Gamma functions as in~\eqref{eq:Gammaexp}, and the mass term as in~\eqref{eq:massexp}. 
Plugging these expressions into the effective actions, and omitting all terms of 
$\mathcal{O}(\epsilon)$, we end up with 
\begin{widetext}
\begin{equation} \label{eq:Seffgen}
    S_{\rm{eff}} = -\frac{1}{8\pi^2} \int d^4x \sqrt{-g} \, 
      \Biggl[ \mathfrak{a}_0 \, \frac{m^4}{4} \, \left( \frac{2}{\epsilon} + \frac{1}{2} 
      -\gamma - 2\ln \frac{\mu}{m} \right) 
      + \mathfrak{a}_1 \, \frac{m^{2}}{2} \, \left( \frac{2}{\epsilon} -\gamma - 
      2 \ln \frac{\mu}{m} \right)  + \mathfrak{a}_2 \, \ln m  + ... \Biggr] 
\end{equation}
\end{widetext}
where 
\begin{eqnarray} \label{eq:coeffgen}
	\mathfrak{a}_j = \begin{cases}
    a_j~~~\,  \quad \quad \quad \quad \quad ~~ \mathrm{scalar~fields} \,,\\[1.1ex]
    - \tilde{a}_j~~~\, \quad \quad \quad \quad ~~ \mathrm{spinor~fields} \,,\\[1.1ex]
    \tr_{\rm{L}}\left[ g^{\mu\lambda} a_{j\lambda\nu}^{~} \right] 
    	\quad~~ \mathrm{vector~fields} \,.\end{cases} 
\end{eqnarray}
In expression~\eqref{eq:Seffgen}, considering $\epsilon \rightarrow 0$, we have now 
clearly isolated the divergent and convergent parts of the effective action, and can 
proceed to the argument of cancelation. The aim 
is to obtain an effective action which does not contain any divergent terms and whose 
convergent remainder is small enough to explain the perceived accelerated expansion of 
the universe. In order to eliminate the divergent contributions to the effective action, 
the masses $m_i$ of the particles in the system have to fulfill the condition 
\begin{equation} \label{eq:div}
  \sum_{i} \left[ \mathfrak{a}_{0,i} \, \frac{m_i^4}{4} + \mathfrak{a}_{1,i}\, \frac{m_i^2}{2} \right] =0 \,.
\end{equation} 
Then the remaining convergent part of the effective action must have exactly the size 
of the observed cosmological constant driving the accelerated expansion, i.e., the 
system of masses must simultaneously obey 
\begin{eqnarray} \label{eq:conv}
	&~& \frac{1}{8\pi^2} \sum_{i} \bigg[- \mathfrak{a}_{0,i} \, \frac{m_i^4}{2} \ln m_i \\
   &~& ~~~ + 
      \mathfrak{a}_{1,i} \, m_i^2 \left( -\frac{1}{4} - \ln m_i \right) 
      + \mathfrak{a}_{2,i} \, \ln m_i \bigg] = \rho_{\Lambda} \,. \nonumber
\end{eqnarray}
These two conditions can, for a system with any number of particles, be fulfilled by the 
introduction of two new masses. The factor of $\mu$ in the log-terms has not been included 
here since these terms can be eliminated by condition~\eqref{eq:div} -- simply by separating 
the log of the fraction into a sum of two logs, the $\ln \mu$--term then represents a constant 
factor which is canceled by the balance of divergences. As a consequence, the terms in~\eqeqref{eq:conv}, 
in particular the terms proportional to $\mathfrak{a}_{0}$ and $\mathfrak{a}_{1}$, changed sign, 
and thus the contributions of bosons to the sub-leading convergent terms of the effective action 
are now positive, and those of fermions are negative. \\
We would like to return shortly to the case of massless particles, in particular 
massless vector bosons like the photon or gluons. From the effective action for vector 
bosons by analogy with~\eqeqref{eq:effaction}, for massless particles in the limit 
$m\rightarrow 0$ we have to solve integrals of the form 
\begin{equation}
	I_{\alpha} (D) = \int \frac{d^Dk}{(2\pi)^D} \, \frac{1}{(k^2)^{\alpha}} \,. 
\end{equation}
However, these integrals are zero for $D, \alpha \in \mathbb{C}$ when using the formalism of 
dimensional regularization, due to Veltman's formula~\cite{1975Leib,2001HKVSFp107}. 
That na\"\i vely implies that the contributions of massless vector bosons, or massless particles in 
general, are zero. However, we know that there are the kinetic energy contributions to the 
effective action even from massless particles, as argued before in~\secref{sec:summingup}, which are 
lost due to the use of dimensional regularization. Thus, 
to compensate for the missing divergences, we recall another condition to be fulfilled, namely 
the balance of degrees of freedom, as already stated in~\eqref{eq:dof}: 
\begin{equation} \label{eq:dof2}
  \sum_{\rm{b}} \nu_{\rm{b}} = \sum_{\rm{f}} \nu_{\rm{f}} \,. 
\end{equation}
In this balance of degrees of freedom, as well as in condition~\eqref{eq:div}, bosons as 
usual give negative contributions to the effective action, and fermions enter with positive signs. Only 
in the sub-leading convergent remainder, the signs change and bosons occur with positive and fermions 
with negative contributions. \\
In summary, we have three conditions to be fulfilled by a system of particles in order 
to reproduce the observed effect of accelerated expansion, i.e.,~\eqsref{eq:div},~\eqref{eq:conv} 
and~\eqref{eq:dof2}. We shall see that it is possible to fulfill these three conditions by the 
introduction of only two new particles to the system.

\section{Dark energy from the curvature terms in the effective action}
\label{sec:DE}
In the previous sections, we have obtained expressions for the vacuum energy 
of a system of bosons and fermions as an expansion in terms of the curvature of the 
system, and set up conditions for the cancelation of the effective action depending 
on the particle masses and the curvature of spacetime. For a FLRW-universe described 
by the metric 
\begin{equation}
	ds^2 = dt^2 - a(t)^2 \left( dx^2 + dy^2 + dz^2 \right)
\end{equation}
we can now calculate the relevant curvature quantities like the curvature scalar, the Ricci tensor 
and the Riemann tensor, and then proceed to compute the coefficients of the 
heat kernel expansion as given by Eqs.\,\eqref{eq:expParker},~\eqref{eq:coeffVectors}, 
and~\eqref{eq:spinorcoeff}. The results can be found in the appendix. They are expressions 
depending on the scale factor and its derivatives, which can be recast as a function of a series 
of cosmological parameters, the so-called cosmographic series (CS). The most prominent of them 
is the Hubble parameter $H_0$, followed by several more, defined as 
\begin{eqnarray} \label{eq:CS}
  H &\equiv & \frac{1}{a} \frac{da}{dt}\,,\quad\quad\quad~~ q \equiv -\frac{1}{aH^2} \frac{d^2a}{dt^2}\,, \\
  j &\equiv & \frac{1}{aH^3} \frac{d^3a}{dt^3}\,, \quad ~~ s \equiv \frac{1}{aH^4} \frac{d^4a}{dt^4}\,. \nonumber
\end{eqnarray}
The parameter $q$ is dubbed acceleration parameter, since it describes the acceleration 
behavior of the universe. Further we have the jerk parameter $j$, capturing inflection points in the kinematic 
history of the universe, and the snap parameter $s$ from the next order of expansion. 
From the analysis of experimental data, in particular the numerical fits of the 
luminosity-redshift relation of supernovae events, it is possible to obtain numerical values 
for these parameters at the current time $t_0$. In~\cite{2012Avil} the best fit for the CS assuming the 
validity of the $\Lambda$CDM model was obtained as 
\begin{eqnarray}
	H_0 &=& 74.05 \,, \quad \quad q_0 = -0.663 \,, \\
	j_0 &=& 1 \,, \quad \quad \quad ~~s_0 = -0.206 \,. \nonumber
\end{eqnarray}
where $H_0$ is given in units of km/s/Mpc, and the parameters $q_0$, $j_0$ and $s_0$ are 
dimensionless. We can thus express the coefficients of the curvature expansion in terms of 
these parameters, with the results to be found in the appendix as well. \\
If we assume the vacuum energy to be the cause of the accelerated expansion of the universe, 
corresponding to a cosmological constant like the one introduced by Einstein into his field 
equations, it occurs in the $\Lambda \rm{CDM}$ action as 
\begin{equation}
  S_{\Lambda \rm{CDM}} = \int d^Dx \sqrt{-g} \left[ \frac{1}{2\kappa} \left( R-2\Lambda 
      \right) + \mathcal{L}_M \right]\,,
\end{equation}
with $\kappa=8\pi G$, and $\mathcal{L}_M$ representing all matter contributions. The energy 
density ascribed to the cosmological constant is then $\rho_{\Lambda}$, defined as 
\begin{equation}
  \rho_{\Lambda}=\Lambda/\kappa \,.
\end{equation}
From observations, the magnitude of this energy density can be inferred~\cite{2006Cope} to be 
\begin{equation}
  \rho_{\Lambda} \simeq 10^{-122}\, \rho_P \simeq 10^{-47}\, \rm{GeV}^4 \,,
\end{equation}
where $\rho_P \equiv m_P^4$ is the Planck density, and we use a unit system with $c=\hbar=1$. This energy 
density has to be reproduced from the contributions of bosons and fermions in the system. 
The cosmological term enters into the action with a negative sign. Thus, 
since according to~\eqeqref{eq:conv} fermions give a negative contribution to the convergent 
remainder of the effective action, we need ultimately a tiny fermionic excess in the vacuum energy. \\
Considering all of the above input, i.e., the cancelation of the degrees of freedom, and the 
balancing conditions in order to achieve cancelation 
of divergences and make the convergent part equal to the observed cosmological constant, 
using the curvature coefficients calculated for the case of a FLRW universe, we can now consider 
a system of particles and investigate how the aim of explaining the accelerated expansion of the 
universe can be accomplished. 
For the particle content, we use the standard model of particle physics, containing one scalar boson, 
nine Dirac fermions, three Majorana fermions and three vector bosons in the massive sector, and further 
the photon and eight gluons as massless vector bosons, as shown in~\tabref{fig:SM}. 
\begin{table}
    \centering
    \begin{tabular}{c|c|c|c}
    \hline \hline 

    \hline 
    Spin  &   name  & mass   & deg. of freed.\\
    \midrule
    \hline \hline 
    
    \hline
    {\small 0}  &   $H$   & $125.3\,\rm{GeV}$ & $1$ \\ 
     \hline
    {\small 1/2}  &   $u$, $\bar{u}$   & $2.4\,\rm{MeV}$ & $4$ \\ 
		      &   $d$, $\bar{d}$   & $4.8\,\rm{MeV}$ & $4$ \\ 
		      &   $c$, $\bar{c}$   & $1.27\,\rm{GeV}$ & $4$ \\ 
		      &   $s$, $\bar{s}$   & $104\,\rm{MeV}$ & $4$ \\ 
		      &   $t$, $\bar{t}$   & $171.2\,\rm{GeV}$ & $4$ \\ 
		      &   $b$, $\bar{b}$   & $4.2\,\rm{GeV}$ & $4$ \\ 
		      &   $e^-$, $e^+$   & $0.511\,\rm{MeV}$ & $4$ \\ 
		      &   $\mu^-$, $\mu^+$   & $105.7\,\rm{MeV}$ & $4$ \\ 
		      &   $\tau^-$, $\tau^+$   & $1.777\,\rm{GeV}$ & $4$ \\ 
		      &   $\nu_e$, $\bar{\nu_e}$   & $< 2.2\,\rm{eV}$ & $2$ \\ 
		      &   $\nu_{\mu}$, $\bar{\nu_{\mu}}$   & $<0.17\,\rm{MeV}$ & $2$ \\ 
		      &   $\nu_{\tau}$, $\bar{\nu_{\tau}}$   & $<15.5\,\rm{MeV}$ & $2$ \\ 
    \hline
    {\small 1}        &   $\gamma$   	& $0$ & $2$ \\ 
		      &   $8\,g$   	& $0$ & $8\cdot 2$ \\ 
		      &   $Z^0$  	& $91.2\,\rm{GeV}$ & $3$ \\ 
		      &   $W^{+}$, $W^{-}$   & $80.4\,\rm{GeV}$ & $2\cdot 3$ \\ 
    \bottomrule 
    \hline \hline
    \end{tabular}

    {\tiny Note. Masses in units where $c=\hbar=1$. 

    }
    \caption{Particles of the standard model.}
    \label{fig:SM}
\end{table}
The resulting effective action is dominated by fermionic contributions, so in order to balance the 
sums and achieve $\rho_{\Lambda}$, new bosonic particles are needed to compensate the 
fermionic divergences as well as to reduce the fermionic excess in the convergent remainder down to 
the small value of the observed vacuum energy. It turns out that considering the masses and the 
curvature coefficients in the chosen unit system of $\rm{GeV}^4$ with $c=\hbar=1$, 
in the balance~\eqref{eq:conv} the third term becomes negligible since its order of magnitude lies 
far below that of $\rho_{\Lambda}$. We can thus simplify~\eqref{eq:conv} to 
\begin{equation} \label{eq:convnew}
   \frac{1}{8\pi^2} \sum_{i} \bigg[- \mathfrak{a}_{0,i} \, \frac{m_i^4}{2} \ln m_i \\
   + \mathfrak{a}_{1,i} \, m_i^2 \left( -\frac{1}{4} - \ln m_i \right) 
   \bigg] = \rho_{\Lambda} \,. 
\end{equation}
Since we have two mass balance equations to be fulfilled, i.e.,~\eqsref{eq:div} and~\eqref{eq:convnew}, 
we need to postulate at least two new bosons. Further, we need to fulfill the balance of the degrees 
of freedom. Due to the abundance of fermions, $42-28=14$ additional bosonic degrees of freedom are required 
to fulfill~\eqeqref{eq:dof2}. This gives constraints on the nature of the new bosons, i.e., on their spin 
and charge. The charges involved are probably none of those known from the standard model, since in that case 
the new particles should participate in the interactions with the standard model particles and be detected in 
experiments. \\
In the case of two new particles, the 14 degrees of freedom can be distributed in a $4+10$ manner, i.e., ten of 
the $14$ degrees of freedom could be due to a spin 2 particle with a $U(1)$ type of charge, while the four 
remaining degrees of freedom could be explained by a spin 1 particle with a more complicated construction of 
charges, possibly similar to the $U(1) \times SU(2)$ group of the electroweak theory. Thus postulating two new 
particles, one vector and one tensor boson, from~\eqsref{eq:div} and~\eqref{eq:convnew} we can calculate their 
masses as 
\begin{eqnarray}
  m_V &=& 168.9 \,\rm{GeV} \,,\\
  m_T &=& 67.7 \,\rm{GeV} \,. \nonumber
\end{eqnarray}
These two new particles with the above mentioned properties can thus reproduce the desired value of 
the vacuum energy corresponding to the acceleration behavior of the universe. The massive tensor 
boson is reminiscent of the as yet unknown quantized form of the gravitational field. \\
Of course, this is only the minimal solution of our conditions in order to achieve cancelation. 
In principle, it is possible to fulfill the conditions with combinations of more than two particles -- 
we are free to play with the balance of the degrees of freedom and the masses of the newly introduced 
particles. Instead of the above construction, the degrees of freedom could also be distributed to 
three new particles in a $3+10+1$ manner, with an uncharged vector boson with mass $m_V$, a charged 
tensor boson with mass $m_T$ and a massless uncharged scalar field $S$ accounting for the last degree of 
freedom. Another possibility would be to consider three massive particles in the balance. However, then 
the system of equations is overdetermined, and one of the three particles can be chosen arbitrarily. 
In one variation of a $3+10+1$ distribution of degrees of freedom, one could consider the axion 
field~\cite{1977Pecc} as the uncharged scalar boson, with its mass generally assumed to be bound by the order 
of $m_a \lesssim 1\, \rm{eV}$, and, as before, two higher spin fields like one uncharged vector and one charged 
tensor. The masses of the vector and tensor bosons change due to the introduction of the axion field, and are 
calculated as 
\begin{eqnarray}
  m_V &=& 57.6 \,\rm{GeV} \,,\\
  m_T &=& 174.3 \,\rm{GeV} \,. \nonumber
\end{eqnarray}
As a last example, we considered a $2+6+6$ distribution of degrees of freedom, assuming a charged scalar 
field and two charged vector fields. Choosing the mass of the charged scalar boson to be again of the order 
of the axion mass, we determined the masses of the two vector bosons as 
\begin{eqnarray}
  m_{V1} &=& 76.9 \,\rm{GeV} \,,\\
  m_{V2} &=& 117.3 \,\rm{GeV} \,. \nonumber
\end{eqnarray}
The different examples proposed here are summarized in~\tabref{fig:Ex}, where the respective type, mass 
and degrees of freedom are listed. \\
Note that in order to achieve the balances, the masses of the particles need to be fine-tuned to a very 
high degree, since the leading terms in the balancing condition are many orders of magnitude higher than 
zero, or the size of the energy density $\rho_{\Lambda}$ that we would like to achieve, respectively. 
The heaviest particles of the standard model, i.e., the Higgs, the top quark and the $W^{\pm}$ bosons, 
dictate the order of magnitude of the masses of the new particles, since they give the most prominent 
contributions to the vacuum energy to be balanced. Thus, in any combination of new fields postulated, at 
least two heavy fields are to be expected. \\

\begin{table}
    \centering
    \begin{tabular}{c|c|c|c}
    \toprule
    \hline \hline 
    
    \hline
    Spin  &   name  & mass   & deg. of freed.\\
    \midrule
    \hline \hline 
    
    \hline
    {\small 1}        &   $V$   & \quad $168.9\,\rm{GeV}$ \quad & $4$ \\ 
    \hline    
    {\small 2}        &   $T$   & $67.7\,\rm{GeV}$  & $10$ \\ 
    \midrule
    \hline \hline 
    
    \hline
    {\small 0}        &   $S$   & $0$               & $1$ \\ 
    \hline
    {\small 1}        &   $V$   & $168.9\,\rm{GeV}$ & $3$ \\ 
    \hline 
    {\small 2}        &   $T$   & $67.7\,\rm{GeV}$  & $10$ \\ 
    \midrule
    \hline \hline 
    
    \hline
    {\small 0}        &   $a$   & $1\,\rm{eV}$ & $1$ \\ 
    \hline 
    {\small 1}        &   $V$   & $174.3\,\rm{GeV}$ & $3$ \\ 
    \hline
    {\small 2}        &   $T$   & $57.6\,\rm{GeV}$  & $10$ \\ 
    \midrule
    \hline \hline
 
    \hline 
    {\small 0}        &  \quad $a^+\,,a^-$ \quad & $1\,\rm{eV}$ & $2$ \\ 
    \hline 
    {\small 1}        &   $V_1$       & $76.8\,\rm{GeV}$ & $2\cdot 3$ \\ 

    ~                 &   $V_2$       & $117.3\,\rm{GeV}$ & $2\cdot 3$ \\ 
    \bottomrule
    \hline \hline
    \end{tabular}

    {\tiny Note. Masses in units where $c=\hbar=1$. 

    }
    \caption{Proposed extensions of the standard model.}
    \label{fig:Ex}
\end{table}

\section{Conclusions}
\label{sec:conclusions}
In this paper, we have addressed the phenomenon of the accelerated re-expansion of the universe, 
i.e., the dark energy conundrum, or, equivalently, the hierarchy problem, dealing with the 
unphysical infinities contributing to the vacuum energy of quantum fields that arise from 
quantum field theory. \\
Usually argued away by renormalization procedures, we have acknowledged 
their physical reality, and succeeded in finding a model in which their very existence is 
necessary and useful to explain the present expansion rate of the universe. The zero-point fluctuations of quantum 
fields have all the desired properties to be an adequate candidate for dark energy, i.e., the 
spatially constant energy density and the repulsive effect driving the acceleration of the 
universe. However, the vacuum energy itself is an infinite or very large quantity, which is usually 
treated with diverse methods of renormalization, and seems unfit to be used as the origin 
of a physical phenomenon. Some theories like supersymmetry have tried to 
approach the problem in a different way, arguing that the vacuum energy contributions could 
result in a finite quantity by means of a balance of contributions, taking into account that 
the zero point energy has different signs for bosonic and fermionic species. 
In this article, we recycle the main idea of supersymmetry, 
i.e., the balance of the vacuum energy by boson and fermion contributions, but restrict ourselves 
to a much  more inornate framework. We computed the vacuum energy for different species within the 
formalism of conventional quantum field theory, and tried to obtain a balance of contributions 
by postulating the existence of new particles, employing the mutual cancelation of bosons and 
fermions in the vacuum energy. We carried out our calculations in flat and curved 
spacetimes, for the latter case using a formalism known as heat kernel expansion of the Greens 
function, i.e., the expansion of the propagator in terms of a power series, with coefficients 
proportional to the curvature of spacetime. It turns out that in order to achieve the correct 
expansion behavior of the universe today, it is necessary to postulate the existence of at least two
new bosonic particles with appropriate properties to fulfill the required $14$ degrees of freedom. Several 
possible examples of standard model extensions have been calculated, resulting in all cases in at least 
two heavy fields with masses of the order of $10 - 10^2\, \rm{GeV}$. Adding those particles to the standard model, 
the vacuum energy obtained from conventional quantum field theories can be tuned exactly to the value of the 
energy density of expansion determined by astrophysical observations. \\
Of course, this solution to the 
dark energy problem is achieved by fine-tuning -- the masses of the additional particles are 
required to be determined highly accurately in order to exactly cancel the infinities and result in the 
correct value of the finite remainder of the vacuum energy. However, it is remarkable that already with 
the introduction of only two new particles, it is in principle possible to give meaning to a seemingly 
nonsensical prediction of modern quantum theories, and at the same time resolve an as yet 
unexplained phenomenon in astrophysics. Moreover, we emphasize that the newly introduced particles 
are of generic nature and have no different properties than other fields of the same type and spin 
known so far, unlike in most other theories of dark energy, where new kinds of fields are 
introduced with very different and exotic properties than the known types of matter. \\
Finally, we would like to remark that despite its success, the model presented in this work is not completely 
realistic, since it assumes only non-interacting free particles, whereas we know that in reality there is an 
abundance of interactions between particles. These interactions of course contribute to the vacuum energy, 
which means that they alter the balance, and thus the result which we obtained. Furthermore, even 
though considering $\Lambda$CDM as the basis of our work, we have not included the cold dark matter component 
into the standard model particle content of our calculations, due to the lack of consolidated knowledge on 
this particle species. We have however considered the axion, or an axion-like particle, in two of our examples. 
In principle, some of the heavy fields postulated in our investigations could well be dark matter candidates as 
well. In this context, our calculations are an indication for the existence of further WIMP-like bosonic dark 
matter fields. \\
Finally it should be mentioned that a cosmological constant is needed for explaining the re-acceleration 
of the universe only if one assumes the cosmos to be completely isotropic and homogeneous. However, 
simulations of the cosmological evolution taking into account the dominance of dark matter in the 
gravitational attraction of the galaxies~\cite{Millenium,Tartu} 
have shown that matter has a fractal distribution in space. Thus there is a strong influence of 
inhomogeneities upon the equation of motion of the size parameter of the Universe, which could 
result in an effect just like the one of a cosmological constant~\cite{Buchert}, i.e., a constant term 
in the Friedmann equation. 
Assuming a non-Gaussian nature of these fluctuations, and using a coordinate space cutoff at the radius of 
the universe to obtain a finite value of the energy, one may expect that the resulting value is indeed 
of the right magnitude~\cite{KleiNew}. Thus, depending on whether the assumption of homogeneity and isotropy 
of the universe actually holds true, the approaches to explain the acceleration of expansion differ 
fundamentally. \\
In the case of a homogeneous and isotropic universe however, we have demonstrated that balancing the infinite 
vacuum energy to result in a finite value is in principle possible by the introduction of only a small number 
of new particles. This should be regarded as the main conclusion here. Further efforts could be dedicated to 
more detailed calculations of the vacuum energy considering the various interactions between the 
particles of the standard model, and the more speculative sector of the particle content of the 
universe. \\

\acknowledgments 
We would like to thank the anonymous referee for useful suggestions 
for a better illustration of our model. The work of C.G. was partly 
supported by the Erasmus Mundus Joint Doctorate Program by Grant Number 
2010--1816 from the EACEA of the European Commission. \\


\appendix

\begin{widetext}
\section{Coefficients for FLRW universe}
\label{sec:app}
These are the coefficients of the Greens function expansions for scalar, spinor and 
vector particles, calculated in the case of a FLRW universe. 
\begin{subequations} \label{eq:FLRW}
\begin{align}
  a_0 &= \frac{\tilde{a}_0}{2} = 1\,, \\
  a_1 &= \tilde{a}_1 = \frac{\left[ a'(t)^2 + a(t) a''(t)\right]}{a(t)^2}\,, \\
  a_2 &= \frac{51 a'(t)^4 - 20 a(t) a'(t)^2 a''(t) + 21 a(t)^2 a''(t)^2 + 6 a(t)^3 a^{(4)}(t)}{30 \,a(t)^4} \,,\\
  \tilde{a}_2 &= \frac{\left[-37 + 5a(t)^2\right] a'(t)^4 + 140 a(t) a'(t)^2 a''(t) + 6 a(t)^2 
	\left[ 3 a''(t)^2 - 2 a(t) a^{(4)}(t) \right]}{240 \,a(t)^4} \,,\\
  \tr_{\rm{L}}\left[ g^{\mu\lambda} a_{0\lambda\nu}^{~} \right] &= 4 \,, \\
  \tr_{\rm{L}}\left[ g^{\mu\lambda} a_{1\lambda\nu}^{~} \right] &= \frac{-\left[ 1+3 a(t)^2 \right] a'(t)^2 + 2 a(t) a''(t)}{a(t)^2} \,, \\
  \tr_{\rm{L}}\left[ g^{\mu\lambda} a_{2\lambda\nu}^{~} \right] &= \frac{3 a'(t)^4 \left[ -17 + 41a(t)^2 \right] + 10a(t)a'(t)^2 a''(t) \left[ 
    14 - 15 a(t)^2 \right]}{30 \,a(t)^4} \nonumber\\
   &~~~ + \frac{a(t)^2 \left\{ -2a''(t)^2\left[ 11+12a(t)^2 \right] + a(t) a^{(4)}(t) \left[ 3+a(t)^2 
    \right] \right\}}{10 \,a(t)^4} \nonumber\\
   &~~~ - a(t)^2 a'(t) a^{(3)}(t) \left[ 1+3a(t)^2 \right] \,.
\end{align}
\end{subequations}
Using the definitions of the parameters of the cosmographic series, we can re-express the 
curvature coefficients in terms of the CS as 
\begin{subequations} \label{eq:coeffCS}
\begin{align}
  a_0 &= \tilde{a}_0 = 1\,, \\
  a_1 &= \tilde{a}_1 = - H_0^2 \left( + 1 - q_0 \right)\,, \\
  a_2 &= \frac{H_0^4}{30} \left(51 + 20 q_0 + 21 q_0^2 + 6 s_0 \right) \,,\\
  \tilde{a}_2 &= \frac{H_0^4}{120} \left( -16 - 70 q_0 + 9 q_0^2 - 6 s_0 \right) \,,\\
  \tr_{\rm{L}}\left[ g^{\mu\lambda} a_{0\lambda\nu}^{~} \right] &= 4 \,, \\
  \tr_{\rm{L}}\left[ g^{\mu\lambda} a_{1\lambda\nu}^{~} \right] &= -2 H_0^2 \left( 2 + q_0 \right) \,,\\
  \tr_{\rm{L}}\left[ g^{\mu\lambda} a_{2\lambda\nu}^{~} \right] &= \frac{H_0^4}{15} \left(36 + 5 q_0 - 69 q_0^2 - 60 j_0  + 6 s_0 \right) \,.
\end{align}
\end{subequations}
\end{widetext}

\end{document}